\documentclass[showpacs]{revtex4} 
\usepackage{epsfig,amsmath,amssymb,graphicx,upgreek,textcomp,upgreek}
\usepackage{natbib}
\usepackage[english]{babel} %

\begin{document}

\vspace{0mm}
\title{Phonon and optical-roton branches of excitations of the Bose system} %
\author{Yu.M. Poluektov}
\email{yuripoluektov@kipt.kharkov.ua (y.poluekt52@gmail.com)} %
\affiliation{National Science Center ``Kharkov Institute of Physics and Technology'', 61108 Kharkov, Ukraine} %
\author{A.A. Soroka} %
\affiliation{National Science Center ``Kharkov Institute of Physics and Technology'', 61108 Kharkov, Ukraine} %

\begin{abstract}
For a system of a large number of Bose particles, a chain of coupled
equations for the averages of field operators is obtained. In the
approximation where only the averages of one field operator and the
averages of products of two operators at zero temperature are taken
into account, there is derived a closed system of dynamic equations.
Taking into account the finite range of the interaction potential
between particles, the spectrum of elementary excitations of a
many-particle Bose system is calculated, and it is shown that it has
two branches: a sound branch and an optical branch with an energy
gap at zero momentum. At high density, both branches are
nonmonotonic and have the roton-like minima. The dispersion of the
phonon part of the spectrum is considered. The performed
calculations and analysis of experiments on neutron scattering allow
to make a statement about the complex structure of the Landau
dispersion curve in the superfluid $^4$He.
\newline%
\end{abstract}
\maketitle

\section{Introduction}\vspace{-0mm} 
The Bose-Einstein condensate in low-density systems of weakly
interacting Bose particles at zero temperature is usually described
by the Gross-Pitaevskii equation with local interaction
\cite{R01,R02}, which is currently widely used to study atomic
condensates created in magnetic and laser traps \cite{R03,R04}. The
Gross-Pitaevskii equation is obtained in the self-consistent field
approximation, in which short-range particle correlations are not
taken into account. In this case the Bose system is characterized by
a coherent state vector \cite{R05}. Meanwhile, the account for pair
correlations  which are significant at small distances turns out to
be important even in low-density systems, since it leads to some
qualitatively new results. Thus, in a rarefied gas of classical
particles the account for pair correlations  makes it possible to
obtain the collision integral in the kinetic equation and,
consequently, all the effects described by the Boltzmann equation
\cite{R06}. The role of pair correlations in an equilibrium Bose
system with a condensate was studied in \cite{R07,R08,R09,R10}.
Dynamic equations with allowance for pair correlations in the case
of local interaction were considered in \cite{R11}.

The purpose of this work is to obtain a system of dynamic equations
with taking into account the finite radius of the interparticle
interaction potential, which manifests itself in the dependence of
the Fourier component of the interaction potential on the wave
vector, and to study the spectrum of elementary excitations. In the
developed theory, in addition to the single-particle anomalous
averages that violate the phase symmetry of the state, the pair
correlations are also taken into account, and the correlations of a
larger number of particles are neglected. Elementary excitations
against the background of a spatially homogeneous equilibrium state
are studied. It is shown that, when accounting for pair
correlations, there are two branches of elementary excitations: one
of them is sound, and the other is optical having an energy gap in
the long wavelength limit.  It is shown that the phonon part of the
spectrum has an anomalous dispersion. The effect of the finite
radius of the interparticle interaction potential on the form of the
spectrum is manifested in the fact that at a sufficiently high
density the behavior of the dispersion curves becomes nonmonotonic
and minima appear on them, similar to the roton minimum in the
excitation spectrum of superfluid helium \cite{R12}. The form of the
spectra of elementary excitations for the cases of high and low
density is calculated. From the point of view of the results
obtained in this work, as well as the results of experiments on
neutron scattering \cite{R13,R14,R15,R16,R17,R18,R19,R20}, the
structure of the Landau quasiparticle spectrum in He II is
discussed. An analysis of the performed calculations and experiments
allows to make a statement about the complex composite structure of
the Landau dispersion curve, in which the region at low momenta
belongs to the phonon branch of excitations and the maxon-roton
region of the spectrum is mainly determined by the optical branch.

\section{Equations for the mean field operators }\vspace{-0mm} %
An arbitrary operator in the Heisenberg representation %
$A(t)=e^{\frac{i}{\hbar}Ht}A(0)e^{-\frac{i}{\hbar}Ht}$ %
obeys the dynamic equation
\begin{equation} \label{01}
\begin{array}{l}
\displaystyle{%
   i\hbar\frac{\partial A}{\partial t}=\big[A,H\big],   %
}
\end{array}
\end{equation}
where the Hamiltonian in the second quantization representation can
be written as a sum of the kinetic energy and pair interaction
energy operators $H=H_1+H_2$, and
\begin{equation} \label{02}
\begin{array}{ccc}
\displaystyle{%
   H_1=\int\!d{\bf r}d{\bf r}'H\big({\bf r},{\bf r}'\big)\Psi^+({\bf r},t)\Psi({\bf r}',t),        %
}\vspace{2mm}\\ %
\displaystyle{%
   H_2=\int\!d{\bf r}d{\bf r}'U\big(|{\bf r}-{\bf r}'|\big)H\big({\bf r},{\bf r}'\big)\Psi^+({\bf r},t)\Psi^+({\bf r}',t)\Psi({\bf r}',t)\Psi({\bf r},t).   %
}%
\end{array}
\end{equation}
Here
\begin{equation} \label{03}
\begin{array}{l}
\displaystyle{%
   H\big({\bf r},{\bf r}'\big)=-\frac{\hbar^2}{2m}\Delta\,\delta({\bf r}-{\bf r}')+\big[U_0({\bf r})-\mu\big]\delta({\bf r}-{\bf r}'),   %
}
\end{array}
\end{equation}
and $m$ is the Bose particle mass, $\mu$ is the chemical potential,
$U_0({\bf r})$ is the energy of a particle in an external field. We
will assume that the interaction potential of particles $U\big(|{\bf r}-{\bf r}'|\big)$ %
depends only on the distance between particles, and the spin of
particles is equal to zero. The field operators $\Psi^+({\bf r},t), \Psi({\bf r},t)$ %
obey the standard commutation relations for Bose particles
\cite{R21}. Let $\langle\Psi\rangle$ be the average value of the
field operator over the vacuum state at zero temperature. Then we
can write the field operator by separating the $c$\,\,-\,number and
operator parts in it:
\begin{equation} \label{04}
\begin{array}{l}
\displaystyle{%
   \Psi=\langle\Psi\rangle+\xi, \qquad  \Psi^+=\langle\Psi\rangle^*+\xi^+.    %
}
\end{array}
\end{equation}
Relations (\ref{04}) are the definition of the overcondensate
operators $\xi, \xi^+$, for which the following obvious conditions
are fulfilled:
\begin{equation} \label{05}
\begin{array}{l}
\displaystyle{%
   \langle\,\xi\,\rangle=\langle\,\xi^+\rangle=0.    %
}
\end{array}
\end{equation}
Here and below, the averaging over the vacuum state is understood in
the sense of the quasiaverages for systems with broken phase
symmetry \cite{R22,R23}. We will take into account both the normal
averages, which are invariant under the phase transformation of
field operators $\Psi\rightarrow\Psi'=e^{i\alpha}\Psi$, and the
anomalous averages where this invariance is broken. We emphasize
that it is precisely the existence of anomalous averages that
entails the property of superfluidity. Let us introduce the notation
for the anomalous average of one field operator:
\begin{equation} \label{06}
\begin{array}{l}
\displaystyle{%
   \eta(r,t)\equiv\big\langle\Psi(r,t)\big\rangle,  \qquad %
   \eta^*(r,t)\equiv\big\langle\Psi^+(r,t)\big\rangle. %
}
\end{array}
\end{equation}
In (\ref{06}) and further, where this does not cause
misunderstanding, we also use the notation $r\equiv{\bf r}$. The
averages of products of several field operators can be expressed in
terms of the averages of products of the operators $\xi, \xi^+$.
Thus, for example, the average of products of two field operators,
taking into account (\ref{05}), can be represented as
\begin{equation} \label{07}
\begin{array}{ccc}
\displaystyle{%
   \big\langle\Psi^+(r)\Psi(r')\big\rangle=\eta^*(r)\,\eta(r')+\big\langle\xi^+(r)\xi(r')\big\rangle, %
}\vspace{2mm}\\ %
\displaystyle{%
  \big\langle\Psi(r)\Psi(r')\big\rangle=\eta(r)\,\eta(r')+\big\langle\xi(r)\xi(r')\big\rangle, %
}\vspace{2mm}\\ %
\displaystyle{%
  \big\langle\Psi^+(r)\Psi^+(r')\big\rangle=\eta^*(r)\,\eta^*(r')+\big\langle\xi^+(r)\xi^+(r')\big\rangle. %
}%
\end{array}
\end{equation}
The average products of a larger number of field operators can be
written similarly. They will also contain the averages of a larger
number of the overcondensate operators of the form
$\big\langle\xi^+(r_1)\xi(r_2)\xi(r_3)\big\rangle$, %
$\big\langle\xi^+(r_2)\xi(r_3)\xi(r_4)\big\rangle$, %
$\big\langle\xi^+(r_1)\xi^+(r_2)\xi(r_3)\xi(r_4)\big\rangle$ %
and so on. Assuming successively in the Heisenberg equation
(\ref{01}) the operator $A$ equal to $\Psi, \Psi^+\Psi, \Psi\Psi,\Psi^+\Psi^+\ldots$, %
after averaging we obtain a coupled infinite chain of equations for the averages %
$\langle\Psi\rangle, \langle\Psi^+\Psi\rangle, \langle\Psi\Psi\rangle, \langle\Psi^+\Psi^+\rangle\ldots$, %
similar to the Bogolyubov chain in the kinetic theory of classical
gases \cite{R06} and the chain of equations for the statistical
operator in quantum theory \cite{R24}.

Thus the equation for the average of one field operator (\ref{06}) has the form %
\begin{equation} \label{08}
\begin{array}{ccc}
\displaystyle{%
   i\hbar\,\frac{\partial \eta(r)}{\partial t}=\int\!H(r,r'')\eta(r'')dr''+\int\!U\big(|r-r''|\big)\big\langle\Psi^+(r'')\Psi(r'')\Psi(r)\big\rangle dr'',   %
}%
\end{array}
\end{equation}
and the equations for the normal and anomalous pair correlations are written as %
\begin{equation} \label{09}
\begin{array}{ccc}
\displaystyle{%
   i\hbar\,\frac{\partial \big\langle\Psi^+(r)\Psi(r')\big\rangle}{\partial t}=\int\!\Big[H(r',r'') \big\langle\Psi^+(r)\Psi(r'')\big\rangle-H^*(r,r'')\big\langle\Psi^+(r'')\Psi(r')\big\rangle\Big]dr''-   %
}\vspace{2mm}\\ %
\displaystyle{\hspace{33mm}%
   -\int\!\Big[U\big(|r-r''|\big)-U\big(|r'-r''|\big)\Big]\big\langle\Psi^+(r)\Psi^+(r'')\Psi(r'')\Psi(r')\big\rangle dr'',   %
}%
\end{array}
\end{equation}\vspace{-2mm}
\begin{equation} \label{10}
\begin{array}{ccc}
\displaystyle{%
   i\hbar\,\frac{\partial \big\langle\Psi(r)\Psi(r')\big\rangle}{\partial t}=\int\!\Big[H(r,r'') \big\langle\Psi(r')\Psi(r'')\big\rangle+H(r',r'')\big\langle\Psi(r)\Psi(r'')\big\rangle\Big]dr''+   %
}\vspace{2mm}\\ %
\displaystyle{\hspace{0mm}%
   +\,U\big(r,r'\big)\big\langle\Psi(r)\Psi(r')\big\rangle +\int\!\Big[U\big(|r-r''|\big)+U\big(|r'-r''|\big)\Big]\big\langle\Psi^+(r'')\Psi(r'')\Psi(r)\Psi(r')\big\rangle dr''.   %
}%
\end{array}
\end{equation}
The average product of three field operators, taking into account
(\ref{04}),\,(\ref{05}), can be represented as
\begin{equation} \label{11}
\begin{array}{ccc}
\displaystyle{%
   \big\langle\Psi^+(r_1)\Psi(r_2)\Psi(r_3)\big\rangle=\eta^*(r_1)\eta(r_2)\eta(r_3)+ \eta^*(r_1)\big\langle\xi(r_2)\xi(r_3)\big\rangle +    %
}\vspace{2mm}\\ %
\displaystyle{\hspace{10mm}%
  +\eta(r_2)\big\langle\xi^+(r_1)\xi(r_3)\big\rangle +\eta(r_3)\big\langle\xi^+(r_1)\xi(r_2)\big\rangle +\big\langle\xi^+(r_1)\xi(r_2)\xi(r_3)\big\rangle.%
}%
\end{array}
\end{equation}
The averages of products of four operators entering into
Eqs.\,(\ref{09}) and (\ref{10}) can be represented in a similar way. For example: %
\begin{equation} \label{12}
\begin{array}{ccc}
\displaystyle{%
   \big\langle\Psi^+(r_1)\Psi^+(r_2)\Psi(r_3)\Psi(r_4)\big\rangle=\eta^*(r_1)\eta^*(r_2)\eta(r_3)\eta(r_4)+     %
}\vspace{2mm}\\ %
\displaystyle{\hspace{0mm}%
  +\,\eta^*(r_1)\eta^*(r_2)\big\langle\xi(r_3)\xi(r_4)\big\rangle +\,\eta^*(r_1)\eta(r_3)\big\langle\xi^+(r_2)\xi(r_4)\big\rangle +\,\eta^*(r_1)\eta(r_4)\big\langle\xi^+(r_2)\xi(r_3)\big\rangle +  %
}\vspace{2mm}\\ %
\displaystyle{\hspace{0mm}%
  +\,\eta^*(r_2)\eta(r_3)\big\langle\xi^+(r_1)\xi(r_4)\big\rangle +\,\eta^*(r_2)\eta(r_4)\big\langle\xi^+(r_1)\xi(r_3)\big\rangle +\,\eta(r_3)\eta(r_4)\big\langle\xi^+(r_1)\xi^+(r_2)\big\rangle +  %
}\vspace{2mm}\\ %
\displaystyle{\hspace{0mm}%
  +\,\eta^*(r_1)\big\langle\xi^+(r_2)\xi(r_3)\xi(r_4)\big\rangle +\,\eta^*(r_2)\big\langle\xi^+(r_1)\xi(r_3)\xi(r_4)\big\rangle +  %
}\vspace{2mm}\\ %
\displaystyle{\hspace{0mm}%
  +\,\eta(r_3)\big\langle\xi^+(r_1)\xi^+(r_2)\xi(r_4)\big\rangle +\,\eta(r_4)\big\langle\xi^+(r_1)\xi^+(r_2)\xi(r_3)\big\rangle +  %
}\vspace{2mm}\\ %
\displaystyle{\hspace{0mm}%
  +\,\big\langle\xi^+(r_1)\xi^+(r_2)\xi(r_3)\xi(r_4)\big\rangle.  %
}%
\end{array}
\end{equation}
In order to obtain a closed system from an infinite chain of coupled
equations, in the same way as in the kinetic theory of gases
\cite{R06}, one should approximate higher correlation functions by
products of lower order correlation functions. In what follows, we
will describe the condensate using the single-particle averages
(\ref{06}) and restrict ourselves to taking into account only the
pairwise correlations of the overcondensate operators introduced by
relations (\ref{04}), defining for this purpose the following
correlation functions
\begin{equation} \label{13}
\begin{array}{ccc}
\displaystyle{%
   g(r,r',t)\equiv\big\langle\xi^+(r,t)\xi(r',t)\big\rangle, \qquad     %
}\vspace{2mm}\\ %
\displaystyle{\hspace{0mm}%
  \tau(r,r',t)\equiv\big\langle\xi(r,t)\xi(r',t)\big\rangle, \quad \tau^*(r,r',t)\equiv\big\langle\xi^+(r,t)\xi^+(r',t)\big\rangle. %
}%
\end{array}
\end{equation}
Functions (\ref{13}) have obvious symmetry properties:
\begin{equation} \label{14}
\begin{array}{ccc}
\displaystyle{\hspace{0mm}%
  g(r,r',t)=g^*(r',r,t), \qquad \tau(r,r',t)=\tau(r',r,t), \qquad \tau^*(r,r',t)=\tau^*(r',r,t). %
}%
\end{array}
\end{equation}
The average products of three overcondensate operators, due to the
property (\ref{05}), cannot be expressed in terms of the pair
correlation functions, so they should be set equal to zero: %
$\big\langle\xi^+(r_1)\xi(r_2)\xi(r_3)\big\rangle\approx 0.$ The
average products of four operators will be approximated using the
products of the pair correlation functions, for example:
\begin{equation} \label{15}
\begin{array}{ccc}
\displaystyle{%
  \big\langle\xi^+(r_1)\xi^+(r_2)\xi(r_3)\xi(r_4)\big\rangle\approx \tau^*(r_1,r_2)\,\tau(r_3,r_4) + g(r_1,r_3)\,g(r_2,r_4) + g(r_1,r_4)\,g(r_2,r_3),     %
}\vspace{2mm}\\ %
\displaystyle{\hspace{0mm}%
  \big\langle\xi^+(r_1)\xi(r_2)\xi(r_3)\xi(r_4)\big\rangle\approx g(r_1,r_2)\,\tau(r_3,r_4) + g(r_1,r_3)\,\tau(r_2,r_4) + g(r_1,r_4)\,\tau(r_2,r_3).   %
}%
\end{array}
\end{equation}
For systems described by Hamiltonians quadratic in field operators,
these relations are exact \cite{R25}. In our case, as noted, we use
this expansion to obtain a closed system of equations. This
approximation is consistent, since it leads to the correct
thermodynamic relations and, probably, it is the better the less
dense is the many-particle system under consideration. When only
pair correlations are taken into account and in the approximation
(\ref{15}), from (\ref{08})\,--\,(\ref{10}) it follows a closed
system of equations for the functions $\eta(r,t),\, g(r,r',t)$ and $\tau(r,r',t)$: %
\begin{equation} \label{16}
\begin{array}{ccc}
\displaystyle{%
   i\hbar\,\frac{\partial \eta(r)}{\partial t}=-\frac{\hbar^2}{2m}\Delta\,\eta(r)+\big[U_0(r)-\mu\big]\eta(r)\,+   %
}\vspace{2mm}\\ %
\displaystyle{\hspace{0mm}%
  +\int\!dr''\,U\big(|r-r''|\big)\Big[\big|\eta(r'')\big|^2\eta(r)+\eta^*(r'')\,\tau(r,r'')+\eta(r'')\,g^*(r,r'')+\eta(r)\,g(r'',r'')\Big],   %
}%
\end{array}
\end{equation}
\begin{equation} \label{17}
\begin{array}{ccc}
\displaystyle{%
   i\hbar\,\frac{\partial \tau(r,r')}{\partial t}= U\big(|r-r'|\big)\eta(r)\eta(r') + U\big(|r-r'|\big)\tau(r,r')\,-  %
}\vspace{2mm}\\ %
\displaystyle{%
  -\frac{\hbar^2}{2m}\big(\Delta+\Delta'\big)\tau(r,r')+\big[U_0(r)+U_0(r')-2\mu\big]\tau(r,r')\,+  %
}\vspace{2mm}\\ %
\displaystyle{\hspace{0mm}%
  +\int\!dr''\,U\big(|r-r''|\big)\Big[\big|\eta(r'')\big|^2\tau(r,r')+\eta(r)\eta^*(r'')\tau(r',r'')+\eta(r)\eta(r'')g(r'',r')\,+   %
}\vspace{2mm}\\ %
\displaystyle{\hspace{30mm}%
  +\,g(r'',r'')\tau(r,r') + g(r'',r')\tau(r,r'')+g(r'',r)\tau(r',r'')\Big] +    %
}\vspace{2mm}\\ %
\displaystyle{\hspace{0mm}%
  +\int\!dr''\,U\big(|r'-r''|\big)\Big[\big|\eta(r'')\big|^2\tau(r,r')+\eta(r')\eta^*(r'')\tau(r,r'')+\eta(r')\eta(r'')g(r'',r)\,+   %
}\vspace{2mm}\\ %
\displaystyle{\hspace{30mm}%
  +\,g(r'',r'')\tau(r,r') + g(r'',r)\tau(r',r'')+g(r'',r')\tau(r,r'')\Big],    %
}%
\end{array}
\end{equation}
\vspace{-3mm}
\begin{equation} \label{18}
\begin{array}{ccc}
\displaystyle{%
   i\hbar\,\frac{\partial g(r,r')}{\partial t}= \frac{\hbar^2}{2m}\big(\Delta-\Delta'\big)g(r,r')-\big[U_0(r)-U_0(r')\big]g(r,r')\,-  %
}\vspace{2mm}\\ %
\displaystyle{\hspace{0mm}%
  -\int\!dr''\,U\big(|r-r''|\big)\Big[\big|\eta(r'')\big|^2g(r,r')+\eta^*(r)\eta(r'')g(r'',r')+\eta^*(r)\eta^*(r'')\tau(r'',r')\,+   %
}\vspace{2mm}\\ %
\displaystyle{\hspace{30mm}%
  +\,\tau^*(r,r'')\tau(r'',r') + g(r,r'')g(r'',r')+g(r,r')g(r'',r'')\Big] +    %
}\vspace{2mm}\\ %
\displaystyle{\hspace{0mm}%
  +\int\!dr''\,U\big(|r'-r''|\big)\Big[\big|\eta(r'')\big|^2g(r,r')+\eta(r')\eta^*(r'')g(r,r'')+\eta(r')\eta(r'')\tau^*(r'',r)\,+   %
}\vspace{2mm}\\ %
\displaystyle{\hspace{30mm}%
  +\,\tau(r',r'')\tau^*(r'',r) + g(r,r'')g(r'',r')+g(r,r')g(r'',r'')\Big].    %
}%
\end{array}
\end{equation}

For this system of equations, there holds the condition of
invariance with respect to the time reversal operation, since along
with the solutions $\eta(r,t),\,\tau(r,r',t),\,g(r,r',t)$ it also
has the solutions $\eta^*(r,-t),\,\tau^*(r,r',-t),\,g^*(r,r',-t)$.
If the pair correlations $\tau(r,r'')$ and $g(r,r'')$ are neglected
in Eq.\,(\ref{16}), then it takes the form of the Gross-Pitaevskii
equation \cite{R01,R02}. Note, however, that the system of equations
(\ref{16})\,--\,(\ref{18}) has no solution, in which only the
function $\eta(r,t)$ is non-zero and both pair correlation functions
$\tau(r,r',t),\,g(r,r',t)$ are equal to zero. This means that in the
presence of the single-particle condensate, the system of
interacting particles also inevitably contains the pair condensate.

The mean of the total particle number operator $N$ is given by the formula %
\begin{equation} \label{19}
\begin{array}{ccc}
\displaystyle{%
   \big\langle N\big\rangle = \int\!\Big[\eta^*(r,t)\eta(r,t)+g(r,r,t)\Big]dr. %
}
\end{array}
\end{equation}
The total particle number density is, obviously, %
$n(r,t)=\big|\eta(r,t)\big|^2+g(r,r,t)$ and the particle number
density in the single-particle condensate is $n_Q(r,t)=\big|\eta(r,t)\big|^2$. %
In the following, where this does not cause misunderstanding, as in
equations (\ref{16})\,--\,(\ref{18}), for brevity we will not
explicitly indicate the dependence of the averages on time.

\section{Transition to quasilocal differential equations }\vspace{-0mm} %
Equations (\ref{16})\,--\,(\ref{18}) are integro-differential. Let
us make some further simplifications. The pair correlation functions
(\ref{13}) depend on two coordinates $r, r'$. It is convenient to
pass to new coordinates $\boldsymbol{\uprho}={\bf r}-{\bf r}'\equiv\rho$ %
and ${\bf R}=\frac{1}{2}({\bf r}+{\bf r}')\equiv R$, then
\begin{equation} \label{20}
\begin{array}{ccc}
\displaystyle{%
   \tau(r,r')=\tau\Big(R+\frac{\rho}{2},R-\frac{\rho}{2}\Big)\equiv\tilde{\tau}(R,\rho), \qquad  %
   g(r,r')=g\Big(R+\frac{\rho}{2},R-\frac{\rho}{2}\Big)\equiv\tilde{g}(R,\rho).                  %
}
\end{array}
\end{equation}
When changing the coordinate of the center of mass of a pair $R$,
these functions change slowly at distances of the order of action of
the interparticle potential $r_0$. They can be represented as
\begin{equation} \label{21}
\begin{array}{ccc}
\displaystyle{%
   \tilde{\tau}(R,\rho)=\sum_{{\bf k}}\tau_{{\bf k}}(R)e^{i{\bf k}\boldsymbol{\uprho}}, \qquad  %
   \tilde{g}(R,\rho)=\sum_{{\bf k}}g_{{\bf k}}(R)e^{i{\bf k}\boldsymbol{\uprho}},              %
}
\end{array}
\end{equation}
by expanding the dependence on the ``fast'' coordinates
$\boldsymbol{\uprho}$ in a Fourier series. Instead of exact
functions $\tilde{\tau}(R,\rho),\,\tilde{g}(R,\rho)$, we will use
the functions averaged over a macroscopic volume $V_0\sim L^3$, where $L\gg r_0$:  %
\begin{equation} \label{22}
\begin{array}{ccc}
\displaystyle{%
   \tau(R)\approx V_0^{-1}\int\!\tilde{\tau}(R,\rho)\,d^3\rho,  \qquad  %
   g(R)\approx V_0^{-1}\int\!\tilde{g}(R,\rho)\,d^3\rho.                %
}%
\end{array}
\end{equation}
This means that in the expansions (\ref{21}) we will take into
account only terms with ${\bf k}=0$ and omit all other terms, which
contain a factor depending on the distance between two points and
rapidly oscillate with increasing $k$. Let us substitute the
expansions (\ref{21}) into equations (\ref{16})\,--\,(\ref{18}) and
retain, in accordance with the chosen approximation, slowly varying
functions with ${\bf k}=0$. As a result, we arrive at the following
system of dynamic equations:
\begin{equation} \label{23}
\begin{array}{ccc}
\displaystyle{%
   i\hbar\,\frac{\partial \eta(r)}{\partial t}=-\frac{\hbar^2}{2m}\Delta\,\eta(r)+\big[U_0(r)-\mu\big]\eta(r)\,+   %
}\vspace{2mm}\\ %
\displaystyle{\hspace{0mm}%
  +\int\!dr''\,U\big(|r-r''|\big)\Big[\big|\eta(r'')\big|^2\eta(r)+\eta^*(r'')\,\tau\Big(\frac{r+r''}{2}\Big)+\eta(r'')\,g^*\Big(\frac{r+r''}{2}\Big)+\eta(r)\,g(r'')\Big],   %
}%
\end{array}
\end{equation}
\vspace{-4mm}
\begin{equation} \label{24}
\begin{array}{ccc}
\displaystyle{%
   i\hbar\,\frac{\partial \tau(r)}{\partial t}= -\frac{\hbar^2}{4m}\Delta\tau(r)+ U(0)\eta^2(r)+\big[U(0)+2U_0(r)-2\mu\big]\tau(r)\,+    %
}\vspace{2mm}\\ %
\displaystyle{\hspace{0mm}%
  +2\int\!dr''\,U\big(|r-r''|\big)\Big[\big|\eta(r'')\big|^2\tau(r)+\eta(r)\eta^*(r'')\tau\Big(\frac{r+r''}{2}\Big)+\eta(r)\eta(r'')g\Big(\frac{r+r''}{2}\Big) +   %
}\vspace{2mm}\\ %
\displaystyle{\hspace{00mm}%
  +\,g(r'')\tau(r) + 2\,g\Big(\frac{r+r''}{2}\Big)\,\tau\Big(\frac{r+r''}{2}\Big)\Big],    %
}%
\end{array}
\end{equation}
\vspace{-4mm}
\begin{equation} \label{25}
\begin{array}{ccc}
\displaystyle{%
   i\hbar\,\frac{\partial g(r)}{\partial t}=  %
   -\int\!dr''\,U\big(|r-r''|\big)\bigg\{\Big[\eta^*(r)\eta(r'')-\eta(r)\eta^*(r'')\Big]g\Big(\frac{r+r''}{2}\Big) +   %
}\vspace{2mm}\\ %
\displaystyle{\hspace{45mm}%
  +\,\eta^*(r)\eta^*(r'')\tau\Big(\frac{r+r''}{2}\Big)-\eta(r)\eta(r'')\tau^*\Big(\frac{r+r''}{2}\Big)\bigg\}.    %
}%
\end{array}
\end{equation}
Here $U(0)$ is the value of the interaction potential at the origin.
The study of this system of equations for the local case, when it
was assumed that $r''\approx r$ in the functions under the integral
and, therefore, the dependence of Fourier component of the
interaction potential on the wave vector was not taken into account,
was carried out in work \cite{R11}.

In the obtained equations an important role is played by the
behavior of the interparticle interaction potential at small
distances. The form of potential here is poorly known. Moreover, in
most model potentials such as, for example, the Lennard-Jones
potential, it is assumed that at small distances they tend to
infinity \cite{R26,R27}. However, there are potentials, such as the
Morse potential and its modifications \cite{R27}, which take on a
finite value at the origin. Note that the use of model potentials
that tend to infinity at small distances leads to significant
difficulties, since such potentials do not have their Fourier image.
Meanwhile, the requirement of "impermeability" of atoms at
arbitrarily high pressures, which is fulfilled in this case, is
unnecessarily stringent, since obviously there must be a pressure at
which the atom will be "crushed" and cease to exist as a separate
structural unit. Therefore, in our opinion, it is physically
justified and natural to use potentials that take on a finite value
at small distances. Note also that quantum chemical calculations
indicate that the potentials at zero tend to have a finite, albeit
large, value \cite{R28,R29}. Since the potential energy of
interaction of atoms at short distances is poorly known, and the
problem of taking into account the short-range correlations in
quantum systems is rather complicated \cite{R30,R31,R32}, then for
specific calculations we will use a relatively simple model
potential which is a modification of the well-known Sutherland
potential \cite{R26,R27}:
\begin{equation} \label{26}
\begin{array}{ccc}
\displaystyle{%
  U({\bf r})=\left\{
               \begin{array}{ll}
                 \hspace{3mm}   I,                             \hspace{17mm}  r<r_0, \vspace{2mm}\\
                 \displaystyle{-I_m\Big(\frac{r_0}{r}\Big)^6}, \hspace{5mm}  r>r_0.
               \end{array} \right.
}%
\end{array}
\end{equation}
This potential contains three parameters: one $r_0$ with length
dimension is the radius of the repulsive core, and two parameters
with energy dimension are the repulsion intensity $I>0$ and the well
depth $I_m>0$. When neglecting the attraction between particles
$I_m=0$ at $r>r_0$, (26) transforms into the model potential of
``semi-transparent sphere'', which was used in similar calculations
earlier \cite{R11}. In addition it is convenient to introduce two
dimensionless parameters:
\begin{equation} \label{27}
\begin{array}{ccc}
\displaystyle{%
   J\equiv\frac{I_m}{I},  \qquad \theta\equiv\frac{I}{\varepsilon_a},  %
}%
\end{array}
\end{equation}
where the characteristic energy $\varepsilon_a\equiv\frac{\hbar^2}{2mr_0^2}$ %
is determined by the mass of an atom and the radius of the repulsive
core. The permissible ranges of change of parameters (\ref{27}): %
$0\leq J< 1$,\, $0<\theta<\infty$. Along with the parameter $J$, we
will also use the parameter
\begin{equation} \label{28}
\begin{array}{ccc}
\displaystyle{%
   b\equiv\frac{1}{1-J},  %
}%
\end{array}
\end{equation}
for which always $b\geq 1$.

\section{Spatially homogeneous state}\vspace{-0mm} %
Let us consider the equilibrium conditions in the spatially
homogeneous state in the absence of an external field $U_0({\bf
r})=0$, when the quantities $\eta({\bf r})=\eta,\,g({\bf r})=g,\,\tau({\bf r})=\tau$  %
do not depend on the coordinates. Equations
(\ref{23})\,--\,(\ref{25}) in this case give rise to a system of algebraic equations %
\begin{equation} \label{29}
\begin{array}{ccc}
\displaystyle{%
   -\mu\eta +U_0\Big(\eta|\eta|^2+\eta^*\tau+2\eta g\Big)=0,  %
}\vspace{-4mm}%
\end{array}
\end{equation}
\begin{equation} \label{30}
\begin{array}{ccc}
\displaystyle{%
   U(0)\eta^2 +\big[U(0)-2\mu\big]\tau +U_0\Big(4|\eta|^2\tau+2\eta^2g+6g\tau \Big)=0,  %
}\vspace{-3mm}%
\end{array}
\end{equation}
\begin{equation} \label{31}
\begin{array}{ccc}
\displaystyle{%
   \eta^{*2}\tau-\eta^2\tau=0. %
}%
\end{array}
\end{equation}
Here $U_0=\int_V\!U({\bf r}')d{\bf r}'$. The quantity $g$ is real
and positive, and from complex quantities we extract the modulus and phase: %
$\eta=\overline{\eta}e^{i\alpha},\, \tau=\overline{\tau}e^{i\beta}$. %
From (\ref{31}) it follows that $\sin (2\alpha-\beta)=0$. %
Thus, there are two possibilities $2\alpha-\beta=0$ and $2\alpha-\beta=\pi$. %
The second possibility should be chosen, since only in this case
equations (\ref{29}) and (\ref{30}) have physically correct
solutions, leading to $\tau=-\overline{\tau}e^{i2\alpha}$. %
As a result, equations (\ref{29}) and (\ref{30}) take the form
\begin{equation} \label{32}
\begin{array}{ccc}
\displaystyle{%
   \overline{\eta}\,\big[-\!\mu +U_0\left(\overline{\eta}^2-\overline{\tau}+2g\right)\big]=0,  %
}\vspace{-3mm}%
\end{array}
\end{equation}
\begin{equation} \label{33}
\begin{array}{ccc}
\displaystyle{%
   U(0)\,\overline{\eta}^2 - \big[U(0)-2\mu + 4U_0\overline{\eta}^2 \big]\overline{\tau} +U_0\big(2\overline{\eta}^2-6\overline{\tau}\big)g=0.  %
}%
\end{array}
\end{equation}
The total density is a sum of the density of number of particles in
the single-particle condensate and the density of number of
particles forming the pair condensate
\begin{equation} \label{34}
\begin{array}{ccc}
\displaystyle{%
   n=\overline{\eta}^2+g.  %
}%
\end{array}
\end{equation}
If the chemical potential is chosen as an independent variable, then
the density must be given as its function $n=n(\mu)$. This
dependence should be obtained from a microscopic calculation, which,
of course, can be performed only approximately. We will assume this
dependence to be known, without specifying its form. In the case
when the density is chosen as an independent variable, one should
consider as given the dependence $\mu=\mu(n)$.

Let us introduce the notation
\begin{equation} \label{35}
\begin{array}{ccc}
\displaystyle{%
   \upsilon\equiv\frac{U(0)}{nU_0}, \qquad w\equiv\frac{\upsilon}{4}+\frac{1}{2}, \qquad z\equiv\frac{\mu}{nU_0}-2,   %
}%
\end{array}
\end{equation}
and also the dimensionless normalized quantities
\begin{equation} \label{36}
\begin{array}{ccc}
\displaystyle{%
   x\equiv\frac{\overline{\eta}^2}{n}, \qquad  y\equiv\frac{\overline{\tau}}{n}.   %
}%
\end{array}
\end{equation}
The parameter $x$ determines the relative density of the
single-particle condensate, so that $0<x\le 1$, and the parameter
$y$ specifies the modulus of the pair anomalous correlation function
normalized to the total density. For the interaction potential
(\ref{26})\, $U(0)=I,\,U_0=I\upsilon_a(1-J)$, where $\upsilon_a\equiv 4\pi r_0^3\big/3$ %
is the ``atomic volume''. In this case $\upsilon=1\big/n\upsilon_a(1-J)\equiv b/\chi$. %
The quantity $\chi\equiv n\upsilon_a$ determines the ratio of the
volume of an atom $\upsilon_a$ to the volume per one atom, and it is
obviously always less than or of the order of unity. For stability
of the system it is required that the attraction be not too strong
and the condition $J<1$ be satisfied, so that $U_0>0$. Obviously,
the more rarefied the system $\chi\equiv n\upsilon_a\ll 1$, the
greater are the parameters $\upsilon$ and $w$. We will assume that
always $\upsilon>1$, so that $w>3/4$. As the density grows, the role
of triple and higher correlations increases \cite{R33} and,
consequently, the accuracy of the used approximation will
deteriorate. However, we will also consider the limit of high
density. In the dimensionless notation, taking into account that %
$g/n=1-x$, the system (\ref{32}),\,(\ref{33}) takes the form
\begin{equation} \label{37}
\begin{array}{ccc}
\displaystyle{%
   x+y+z=0,  %
}\vspace{-4mm}%
\end{array}
\end{equation}
\begin{equation} \label{38}
\begin{array}{ccc}
\displaystyle{%
   z^2+2(x-w)z+2x^2-4wx=0.  %
}%
\end{array}
\end{equation}
Since the quantities $x$ and $y$ are positive, it follows from
(\ref{37}) that the parameter $z$ must be negative. The quantities
$z=\mu\big/nU_0-2$ and $w=1/2+b/4\chi$\, entering into equations are
determined by the total density of the system, and also the chemical
potential as a function of the density should be found from a
microscopic calculation. Thus, equations (\ref{37}),\,(\ref{38})
allow to determine the dependences of the density of the
single-particle condensate $\overline{\eta}^2=nx$ and the modulus of
the pair anomalous correlation function $\overline{\tau}=ny$ on the
total density, provided that the dependence $\mu=\mu(n)$ is given.
Since such a dependence is not known actually, it is more convenient
to consider as an independent variable the normalized density of the
single-particle condensate, varying within the limits $0<x\le 1$. %
In systems described by the Gross-Pitaevskii equation, it is assumed
that the total density coincides with the density of the
single-particle condensate and, therefore, $x=1$. In superfluid
helium, as is known from experiments on neutron scattering
\cite{R20,R34}, the single-particle condensate constitutes
approximately 10\% of the total density and, therefore, here $x\approx 0.1$. %
Then equations (\ref{37}),\,(\ref{38}) allow to find the normalized
modulus of the pair anomalous correlation function $y$ and the parameter $z$:\vspace{-2mm} %
\begin{equation} \label{39}
\begin{array}{ccc}
\displaystyle{%
   y=\sqrt{w^2+2wx-x^2}-w,  %
}\vspace{-4mm}%
\end{array}
\end{equation}
\begin{equation} \label{40}
\begin{array}{ccc}
\displaystyle{%
   z=-\sqrt{w^2+2wx-x^2}+w-x.  %
}%
\end{array}
\end{equation}
These dependencies are shown in Fig.\,1. As seen from Fig.\,1{\it a}, %
the modulus of the pair anomalous correlation function $\tau$
increases with the density of the single-particle condensate. In a
rarefied system at $w\gg 1$ these values practically coincide $y\approx x$. %
The negative parameter $z$ decreases with increasing the density of
the single-particle condensate (Fig.\,1{\it b}). For it the
inequality $z>-2$ is fulfilled at an arbitrary total density, so the
chemical potential $\mu=nU_0(z+2)$ in the model under consideration
is always positive.
\vspace{0mm} %
\begin{figure}[h!]
\vspace{-0mm}  \hspace{0mm}
\includegraphics[width = 15cm]{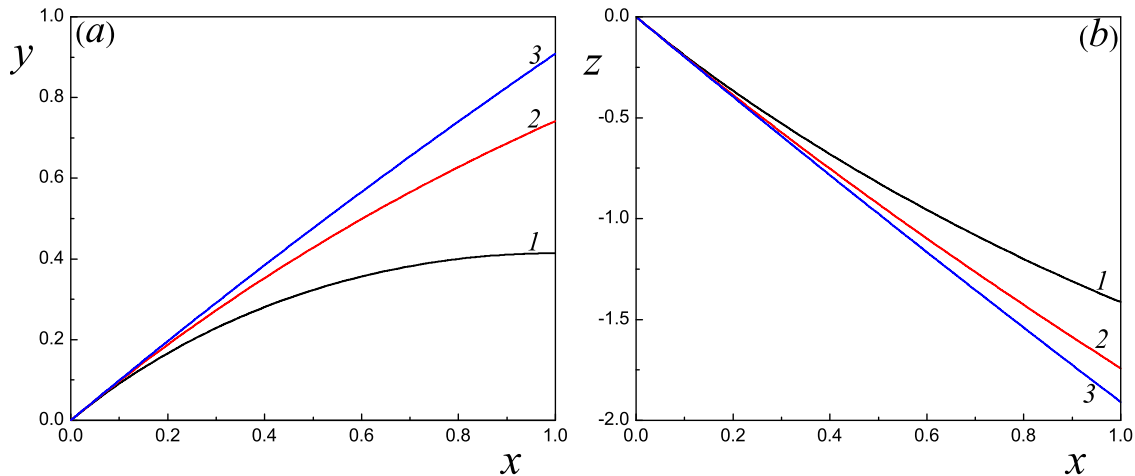} 
\vspace{-3mm} %
\caption{\label{fig01} 
Dependencies of: (\!{\it a}) the modulus of the pair anomalous
correlation function $y=\overline{\tau}/n$  and %
(\!{\it b}) the parameter $z$, determining the chemical potential $\mu=nU_0(z+2)$, %
on the relative density of the single-particle condensate
$x=\overline{\eta}^2\big/n$ at: ({\it 1})~$w=1$ , ({\it 2}) $w=3$, ({\it 3}) $w=10$. %
}%
\end{figure}

We note also that in the presence of the Bose-Einstein condensate
the dependence of macroscopic quantities, in particular energy, on
the magnitude of interaction is, generally speaking, non-analytical.
Therefore, the passage to the limit $U_0\rightarrow 0$  on the
interaction constant in such systems is incorrect. This issue is
discussed in more detail in work \cite{R35}.

\section{Spectrum of elementary excitations}\vspace{-0mm} %
Let us consider the propagation of small perturbations in a
spatially homogeneous system. Assuming
\begin{equation} \label{41}
\begin{array}{ccc}
\displaystyle{%
  \eta(r,t)=\overline{\eta}+\delta\eta(r,t), \qquad  \tau(r,t)=\overline{\tau}+\delta\tau(r,t), \qquad  g(r,t)=g+\delta g(r,t),%
}%
\end{array}
\end{equation}
and denoting for convenience $\delta\varphi(r,t)\equiv\overline{\eta}\delta\eta(r,t)$, %
with an appropriate choice of phases of the complex functions in the
equilibrium state from (\ref{23})\,--\,(\ref{25}) we can obtain a
system of linearized equations for the fluctuations %
$\delta\varphi(r,t),\,\delta\tau(r,t),\,\delta g(r,t)$. It is more
convenient, however, to pass from the complex quantities %
$\delta\varphi(r,t),\,\delta\tau(r,t)$ to real variables:
\begin{equation} \label{42}
\begin{array}{ccc}
\displaystyle{%
  \delta\Psi(r,t)=\delta\varphi(r,t)+\delta\varphi^*(r,t), \qquad  \delta\Phi(r,t)=i\big[\delta\varphi(r,t)-\delta\varphi^*(r,t)\big],  %
}\vspace{2mm}\\ %
\displaystyle{\hspace{0mm}%
  \delta\Theta(r,t)=\delta\tau(r,t)+\delta\tau^*(r,t), \qquad  \delta\Lambda(r,t)=i\big[\delta\tau(r,t)-\delta\tau^*(r,t)\big].  %
}%
\end{array}
\end{equation}
Fluctuations of the particle number density in the single-particle
condensate $\delta n_Q$ and the total particle number density
$\delta n$ are given by the expressions:
\begin{equation} \label{43}
\begin{array}{ccc}
\displaystyle{%
  \delta n_Q(r,t)=\delta\Psi(r,t), \qquad  \delta n(r,t)=\delta\Psi(r,t)+\delta g(r,t).  %
}%
\end{array}
\end{equation}

In the real variables (\ref{42}) the system of linearized equations
takes the form
\begin{equation} \label{44}
\begin{array}{ccc}
\displaystyle{%
  \hbar\,\delta\dot{\Psi}(r)=\frac{\hbar^2}{2m}\Delta\delta\Phi(r)-U_0(\overline{\tau}-g)\delta\Phi(r)-  %
  (g+\overline{\tau})\int\!dr''U\big(|r-r''|\big)\delta\Phi(r'')-   %
}\vspace{2mm}\\ %
\displaystyle{\hspace{0mm}%
  -\,\overline{\eta}^2\int\!dr''U\big(|r-r''|\big)\delta\Lambda\Big(\frac{r+r''}{2}\Big),  %
}%
\end{array}
\end{equation}
\vspace{-4mm} %
\begin{equation} \label{45}
\begin{array}{ccc}
\displaystyle{%
  \hbar\,\delta\dot{\Phi}(r)=-\frac{\hbar^2}{2m}\Delta\delta\Psi(r)+U_0(\overline{\tau}-g)\delta\Psi(r)\,+  %
}\vspace{2mm}\\ %
\displaystyle{\hspace{0mm}%
  +\big(2\overline{\eta}^2+g-\overline{\tau})\int\!dr''U\big(|r-r''|\big)\delta\Psi(r'') %
  +\,\overline{\eta}^2\int\!dr''U\big(|r-r''|\big)\delta\Theta\Big(\frac{r+r''}{2}\Big)\,+  %
}\vspace{2mm}\\ %
\displaystyle{\hspace{0mm}%
  +\,2\overline{\eta}^2\int\!dr''U\big(|r-r''|\big)\delta g \Big(\frac{r+r''}{2}\Big)  %
  +2\overline{\eta}^2\int\!dr''U\big(|r-r''|\big)\delta g\big(r''\big),                %
}
\end{array}
\end{equation}
\vspace{-4mm} %
\begin{equation} \label{46}
\begin{array}{ccc}
\displaystyle{\hspace{0mm}%
  \hbar\,\delta \dot{g}(r)=\int\!dr''U\big(|r-r''|\big)  %
  \bigg\{g\big[\delta\Phi(r'')-\delta\Phi(r)\big]+\overline{\eta}^2\delta\Lambda\Big(\frac{r+r''}{2}\Big)+ %
  \overline{\tau}\delta\Phi(r)+\overline{\tau}\delta\Phi(r'') \bigg\},               %
}
\end{array}
\end{equation}
\vspace{-4mm} %
\begin{equation} \label{47}
\begin{array}{ccc}
\displaystyle{%
  \hbar\,\delta\dot{\Lambda}(r)=-\frac{\hbar^2}{4m}\Delta\delta\Theta(r)\,+  %
}\vspace{2mm}\\ %
\displaystyle{\hspace{0mm}%
  +\big[U(0)+2U_0(\overline{\tau}-g)\big]\delta\Theta(r)+2\big(\overline{\eta}^2+2g\big)\int\!dr''U\big(|r-r''|\big)\delta\Theta\Big(\frac{r+r''}{2}\Big)\,+ %
}\vspace{2mm}\\ %
\displaystyle{\hspace{0mm}%
  +2\big[U(0)+U_0(-\overline{\tau}+g)\big]\delta\Psi(r)+2\big(-3\overline{\tau}+g\big)\int\!dr''U\big(|r-r''|\big)\delta\Psi\big(r''\big)\,- %
}\vspace{2mm}\\ %
\displaystyle{\hspace{0mm}%
  -4\tau\int\!dr''U\big(|r-r''|\big)\delta g\big(r''\big)+4\big(\overline{\eta}^2-2\overline{\tau}\big)\int\!dr''U\big(|r-r''|\big)\delta g\Big(\frac{r+r''}{2}\Big), %
}%
\end{array}
\end{equation}
\vspace{-4mm} %
\begin{equation} \label{48}
\begin{array}{ccc}
\displaystyle{%
  \hbar\,\delta\dot{\Theta}(r)=\frac{\hbar^2}{4m}\Delta\delta\Lambda(r)-2\big[U(0)+U_0(-\overline{\tau}+g)\big]\delta\Phi(r)\,-  %
}\vspace{2mm}\\ %
\displaystyle{\hspace{0mm}%
  -2\big(-\overline{\tau}+g)\int\!dr''U\big(|r-r''|\big)\delta\Phi(r'') %
  -4\overline{\tau}\int\!dr''U\big(|r-r''|\big)\delta\Phi(r'')\,- %
}\vspace{2mm}\\ 
\displaystyle{\hspace{0mm}%
  -\big[U(0)+2U_0(\overline{\tau}-g)\big]\delta\Lambda(r)  %
  -2\big(\overline{\eta}^2+2g\big)\int\!dr''U\big(|r-r''|\big)\delta\Lambda\Big(\frac{r+r''}{2}\Big).               %
}
\end{array}
\end{equation}
We assume that the dependences of fluctuations on coordinates and time have the form %
$\sim\exp\!\!\big[i({\bf k}{\bf r}-\omega t)\big]$ and represent the
Fourier component of the interaction potential %
$U_k=\!\int d{\bf r}U(|{\bf r}|)e^{i{\bf k}{\bf r}}$ %
in the form $U_k=U_0+\Delta U_k$, separating the part %
$\Delta U_k\equiv\!\int d{\bf r}U(|{\bf r}|)\big(e^{i{\bf k}{\bf r}}-1\big)$ %
that depends on the wave vector. Then, taking into account the
notations (\ref{35}),\,(\ref{36}), we arrive at a system of
homogeneous linear algebraic equations:
\begin{equation} \label{49}
\begin{array}{ccc}
\displaystyle{%
  i\hbar\,\omega\delta\Phi=-\big[\varepsilon_k+n\Delta U_k(1+z+2x)+2xnU_0\big]\delta\Psi\,-   %
}\vspace{3mm}\\ %
\displaystyle{\hspace{0mm}%
  -\,xn(U_0+\Delta U_{k/2})\delta\Theta -2xn\big[2U_0+(\Delta U_k+\Delta U_{k/2})\big]\delta g, %
}%
\end{array}
\end{equation}
\vspace{-3mm} %
\begin{equation} \label{50}
\begin{array}{ccc}
\displaystyle{%
  i\hbar\,\omega\delta\Lambda=-\bigg[\frac{\varepsilon_k}{2}+2n\Delta U_{k/2}(2-x)+2nU_0(2w-z-x)\bigg]\delta\Theta\,-   %
}\vspace{3mm}\\ %
\displaystyle{\hspace{0mm}%
  -2\big[2nU_0(2w+2z+x)+n\Delta U_k(2x+3z+1)\big]\delta\Psi\,- %
}\vspace{3mm}\\ %
\displaystyle{\hspace{0mm}%
  -4\big[nU_0(4x+3z)+n\Delta U_k(x+z)+n\Delta U_{k/2}(3x+2z)\big]\delta g, %
}%
\end{array}
\end{equation}
\vspace{-3mm} %
\begin{equation} \label{51}
\begin{array}{ccc}
\displaystyle{%
  i\hbar\,\omega\delta\Psi=\big[\varepsilon_k+n\Delta U_{k}(1-z-2x)-2nU_0(z+x)\big]\delta\Phi + xn\big(U_0+\Delta U_{k/2}\big)\delta\Lambda,   %
}%
\end{array}
\end{equation}
\vspace{-3mm} %
\begin{equation} \label{52}
\begin{array}{ccc}
\displaystyle{%
  i\hbar\,\omega\delta\Theta=\bigg[\frac{\varepsilon_k}{2}+2n\Delta U_{k/2}(2-x)+2nU_0(2w-z-x)\bigg]\delta\Lambda\,+   %
}\vspace{3mm}\\ %
\displaystyle{\hspace{0mm}%
  +2\big[2nU_0(2w-x)+n\Delta U_k(1-2x-z)\big]\delta\Phi, %
}%
\end{array}
\end{equation}
\vspace{-3mm} %
\begin{equation} \label{53}
\begin{array}{ccc}
\displaystyle{%
  i\hbar\,\omega\delta g=\big[2nU_0(x+z)-n\Delta U_{k}(1-2x-z)\big]\delta\Phi - xn\big(U_0+\Delta U_{k/2}\big)\delta\Lambda.   %
}%
\end{array}
\end{equation}
Here $\varepsilon_k=\hbar^2k^2\big/2m$ is the energy of a free
particle. The fluctuations $\delta\Psi,\,\delta\Theta,\,\delta g$ %
in (\ref{51})\,--\,(\ref{53}) can be expressed in terms of the
fluctuations of quantities $\delta\Phi$ and $\delta\Lambda$, for
which we obtain a system of linear equations
\begin{equation} \label{54}
\begin{array}{ccc}
\displaystyle{%
  (\hbar\omega)^2\delta\Phi=A_k\delta\Phi+B_k\delta\Lambda,   %
}\vspace{3mm}\\ %
\displaystyle{\hspace{0mm}%
  (\hbar\omega)^2\delta\Lambda=C_k\delta\Lambda+D_k\delta\Phi.   %
}%
\end{array}
\end{equation}
The coefficients $A_k,\,B_k,\,C_k,\,D_k$ entering here have a rather
cumbersome form, and they are given in Appendix A. From the system
of linear equations (\ref{54}) there follows the dispersion equation
\begin{equation} \label{55}
\begin{array}{ccc}
\displaystyle{%
  (\hbar\omega)^4-L_k(\hbar\omega)^2+N_k=0,    %
}%
\end{array}
\end{equation}
where $L_k\equiv A_k+C_k$ and $N_k\equiv A_kC_k-B_kD_k$. %
From (\ref{55}) we find that there are two branches of elementary excitations %
\begin{equation} \label{56}
\begin{array}{ccc}
\displaystyle{%
  \hbar\omega_k^{(\pm)}=\sqrt{\frac{1}{2}\left(L_k\pm\sqrt{L_k^2-4N_k}\right)}.    %
}%
\end{array}
\end{equation}
Since, as the calculation shows, $N_0=0$, the solution $\omega_k^{(-)}$ %
describes sound excitations, whose energy tends linearly in $k$ to
zero in the long wavelength limit. The solution $\omega_k^{(+)}$
describes optical excitations, whose energy $\hbar\omega_0^{(+)}=\sqrt{L_0}$ %
is finite at $k=0$. Note that the propagation of excitations of both
types is accompanied by the density fluctuations, so that they can
be detected, as it actually takes place in reality, in neutron
scattering experiments \cite{R13,R14,R15,R16,R17,R18,R19,R20}. %

In order for the excitations to be undamped, the obvious condition
must hold $L_k^2-4N_k\geq 0$. If also $L_k>0$, then the excitations
of the optical branch are undamped for all $k$, and the excitations
of the sound branch remain undamped when the condition $N_k\geq 0$ %
holds. In what follows, we will consider only the case $L_k>0$, as
it takes place in the system under consideration.

\section{Dispersion curves for the modified Sutherland potential}\vspace{-0mm} %
Let us analyze the form of dispersion curves in case of interaction
of particles with the potential (\ref{26}). In this case
\begin{equation} \label{59}
\begin{array}{ccc}
\displaystyle{%
  U_0=\upsilon_aI(1-J),  %
}\vspace{3mm}\\ %
\displaystyle{\hspace{0mm}%
  \Delta U_k\equiv U_k-U_0\equiv 3\upsilon_aIf(\kappa),   %
}%
\end{array}
\end{equation}
where the function
\begin{equation} \label{60}
\begin{array}{ccc}
\displaystyle{%
  f(\kappa)\equiv g(\kappa)-\frac{J}{4}\,q(\kappa)  %
}%
\end{array}
\end{equation}
is expressed through the functions
\begin{equation} \label{61}
\begin{array}{ccc}
\displaystyle{%
  g(\kappa)\equiv\frac{(\sin\kappa-\kappa\cos\kappa)}{\kappa^3}-\frac{1}{3},  %
}\vspace{3mm}\\ %
\displaystyle{\hspace{0mm}%
  q(\kappa)\equiv\frac{\sin\kappa}{\kappa}+\frac{1}{3}\cos\kappa-\frac{\kappa}{6}\sin\kappa  %
  -\frac{\kappa^2}{6}\cos\kappa+\frac{\kappa^3}{6}\bigg(\frac{\pi}{2}-\int_0^\kappa\frac{\sin y}{y}\,dy\bigg)-\frac{4}{3}. %
}%
\end{array}
\end{equation}
Here the notation $\kappa\equiv kr_0$ is introduced. The function
$q(\kappa)$ describes the influence of the attractive part of the
potential on the shape of dispersion curves. At $\kappa=0$ the
functions (\ref{60}),\,(\ref{61}) turn to zero, and at small wave
numbers $\kappa\ll 1$:
\begin{equation} \label{62}
\begin{array}{ccc}
\displaystyle{%
  g(\kappa)\approx -\frac{\kappa^2}{30}, \qquad q(\kappa)\approx -\frac{2}{3}\,\kappa^2, \qquad  %
  f(\kappa)\approx\frac{1}{6}\left(J-\frac{1}{5}\right)\!\kappa^2. %
}%
\end{array}
\end{equation}
As we can see, there is a critical value of the parameter $J_*=1/5$\, %
at which the function $f(\kappa)$ changes sign at small values of $\kappa$. %
In the opposite case $\kappa\gg 1$:
\begin{equation} \label{63}
\begin{array}{ccc}
\displaystyle{%
  g(\kappa)= -\frac{1}{3}, \qquad q(\kappa)= -\frac{4}{3}, \qquad  %
  f(\kappa)=-\frac{1}{3}(1-J)\equiv -\frac{1}{3b}. %
}%
\end{array}
\end{equation}
In this limit the optical branch goes over to the dispersion law of
a free particle, and the sound branch to the dispersion law of a
particle with a doubled mass:
\begin{equation} \label{64}
\begin{array}{ccc}
\displaystyle{%
  \hbar\omega_k^{(+)}=\frac{\hbar^2k^2}{2m}, \qquad \hbar\omega_k^{(-)}=\frac{\hbar^2k^2}{4m}.   %
}%
\end{array}
\end{equation}

In numerical calculations it is more convenient to use the
dimensionless form of coefficients of the dispersion equation:
\begin{equation} \label{65}
\begin{array}{ccc}
\displaystyle{%
  \tilde{A}_k\equiv A_k\big/(nU_0)^2, \quad \tilde{B}_k\equiv B_k\big/(nU_0)^2, \quad    %
  \tilde{C}_k\equiv C_k\big/(nU_0)^2, \quad \tilde{D}_k\equiv D_k\big/(nU_0)^2,    %
}\vspace{3mm}\\ %
\displaystyle{\hspace{0mm}%
  \tilde{L}_k\equiv L_k\big/(nU_0)^2=\tilde{A}_k+\tilde{C}_k,  \quad  %
  \tilde{N}_k\equiv N_k\big/(nU_0)^4=\tilde{A}_k\tilde{C}_k-\tilde{B}_k\tilde{D}_k. %
}%
\end{array}
\end{equation}
Explicit expressions of the coefficients of the dispersion equation
(\ref{55}) for the potential (\ref{26}) are given in Appendix B. In
terms of dimensionless quantities the dispersion laws (\ref{56}) can
be written as
\begin{equation} \label{66}
\begin{array}{ccc}
\displaystyle{%
  \varepsilon_k^{(\pm)}\equiv\frac{\hbar\omega_k^{(\pm)}}{nU_0}=\sqrt{\frac{1}{2}\left(\tilde{L}_k\pm\sqrt{\tilde{L}_k^2-4\tilde{N}_k}\right)}.    %
}%
\end{array}
\end{equation}

In the long wavelength limit $\kappa\ll 1$, the expansions are valid: %
$\tilde{A}_k\approx \tilde{A}_0+\tilde{A}_2\kappa^2$,\,
$\tilde{B}_k\approx \tilde{B}_0+\tilde{B}_2\kappa^2$,\,
$\tilde{C}_k\approx \tilde{C}_0+\tilde{C}_2\kappa^2$,\,
$\tilde{D}_k\approx \tilde{D}_0+\tilde{D}_2\kappa^2$. %
The coefficients of these expansions are given in Appendix C. In
this limit the dispersion laws for the sound and optical branches
take the form\vspace{-1mm}
\begin{equation} \label{67}
\begin{array}{ccc}
\displaystyle{%
  \omega_k^{(-)}=ck, \qquad \omega_k^{(+)}=\omega_0+\alpha k^2,   %
} \vspace{-2mm} %
\end{array}
\end{equation}
or\vspace{-2mm}
\begin{equation} \label{68}
\begin{array}{ccc}
\displaystyle{%
  \varepsilon_k^{(-)}=\tilde{c}\kappa, \qquad \varepsilon_k^{(+)}=\varepsilon_0+\tilde{\alpha} \kappa^2,   %
}%
\end{array}
\end{equation}
where $c$ is the speed of sound, $\omega_0$ is the gap in the
optical branch. The corresponding dimensionless parameters in
(\ref{68}) are:
\begin{equation} \label{69}
\begin{array}{ccc}
\displaystyle{%
  \tilde{c}\equiv\frac{\hbar}{nU_0r_0}\,c,       \qquad  %
  \varepsilon_0\equiv\frac{\hbar\omega_0}{nU_0}, \qquad  %
  \tilde{\alpha}\equiv\frac{\hbar}{nU_0r_0^2}\,\alpha.   %
}%
\end{array}
\end{equation}
With allowance for introduced notations of dimensionless quantities,
the speed of sound, the frequency of homogeneous oscillations and
the coefficient in the optical branch are determined by the formulas: %
\begin{equation} \label{70}
\begin{array}{ccc}
\displaystyle{%
  \tilde{c}=\sqrt{\frac{\tilde{A}_0\tilde{C}_2+\tilde{C}_0\tilde{A}_2-\tilde{B}_0\tilde{D}_2-\tilde{D}_0\tilde{B}_2}{\tilde{A}_0+\tilde{C}_0}}, \qquad    %
  \varepsilon_0=\sqrt{\tilde{A}_0+\tilde{C}_0},   %
}\vspace{3mm}\\ %
\displaystyle{\hspace{0mm}%
  \tilde{\alpha}=\frac{\big(\tilde{A}_0\tilde{A}_2+\tilde{C}_0\tilde{C}_2+\tilde{B}_0\tilde{D}_2+\tilde{D}_0\tilde{B}_2\big)}{2\big(\tilde{A}_0+\tilde{C}_0\big)^{3/2}}.  %
}%
\end{array}
\end{equation}
In numerical calculations we will fix the parameters of the
potential: the radius of the repulsive core $r_0$, the parameters
$J$ or $b=1/(1-J)$, and the parameter $\theta\equiv I/\varepsilon_a$
determining the intensity of particle repulsion at small distances.
The system density $n$ is assumed to be known. Then %
$\displaystyle{nU_0=\varepsilon_a\frac{\chi\theta}{b}}$, where recall that %
$\displaystyle{\varepsilon_a\equiv\frac{\hbar^2}{2mr_0^2}}$\, and $\chi\equiv n\upsilon_a$. %

{\it Dense system}. Consider first a dense system with parameters
close to those of liquid helium-4: %
$m=6.65\cdot\!10^{-24}$\,g,\, $n=2.2\cdot\!10^{22}$\,cm$^{-3}$. %
We take the radius of the core $r_0=2.24\cdot\!10^{-8}$\,cm, %
so that in this case $\chi\approx 1$ and $\varepsilon_a=1.67\cdot\!10^{-16}$\,erg$\,\,\approx 1.2$\,K. %
Other parameters of the potential (\ref{26}) are set as follows: %
$\theta=100$\, and $J=0.1$. In accordance with the experimental data
on neutron scattering \cite{R20,R34}, we consider $x\approx 0.1$.
With the chosen fitting parameters, we obtain the following
numerical values of the sound speed, the frequency of homogeneous
oscillations and the coefficient in the optical branch:
\begin{equation} \nonumber
\begin{array}{ccc}
\displaystyle{%
  c=4.3\cdot\!10^{4}\,{\rm cm}\!\cdot{\rm s}^{-1},          \qquad  %
  \frac{\omega_0}{2\pi}=7.8\cdot\!10^{12}\,{\rm s}^{-1},    \qquad  %
  \alpha=-3.7\cdot\!10^{-4}\,{\rm cm}^{2}\!\cdot{\rm s}^{-1}.       %
}%
\end{array}
\end{equation}
The used set of parameters leads to the velocity of phonons that
somewhat exceeds the velocity of the low-frequency hydrodynamic
first sound in superfluid helium
$c_1=2.3\cdot\!10^{4}\,{\rm cm}\!\cdot{\rm s}^{-1}$. %
The numerically calculated dispersion curves in a superfluid liquid
are shown in Fig.\,2.
\begin{figure}[t!]
\vspace{-2mm}  \hspace{0mm}
\includegraphics[width = 7.02cm]{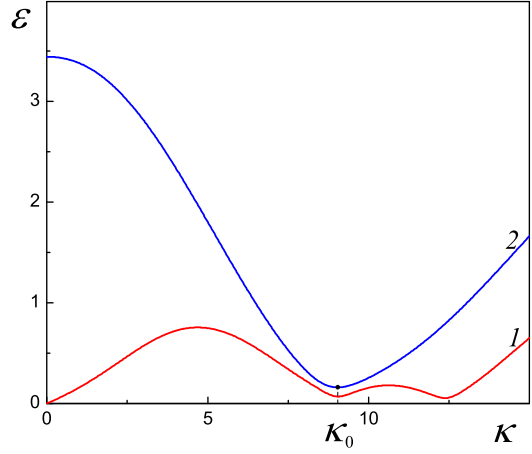} 
\vspace{-3mm} %
\caption{\label{fig02} 
Phonon (\!{\it 1}) $\varepsilon=\big(\hbar\omega^{(-)}\!\big/nU_0\big)$ %
and optical (\!{\it 2}) $\varepsilon=\big(\hbar\omega^{(+)}\!\big/nU_0\big)$ %
branches in the dense system with $n=2.2\cdot\!10^{22}$\,cm$^{-3}$ and parameters: %
$m=6.65\cdot\!10^{-24}$\,g,\,\,$r_0=2.24\cdot\!10^{-8}$\,cm,\,\,$x=0.1$,\,\,$\chi=1$,\,\,$\theta=100$,\,\,$J=0.1$. %
}%
\vspace{-2mm} %
\end{figure}

The lower curve describes the sound excitations in the long
wavelength limit and it has the roton-like minimum at a certain
value of the wave number. Within the framework of the nonlocal
Gross-Pitaevskii equation, in the case when pair correlations are
not taken into account, a similar nonmonotonic curve was obtained in
\cite{R36}. The appearance of the upper curve with a gap in the
limit $k\rightarrow 0$\, is a consequence of taking into account
pair correlations, and it also has the roton-like minimum at
$\kappa_0$ close to the minimum on the sound branch.

{\it Rarefied system}. With a decrease in the density and,
consequently, with an increase in the average distance between
particles, the effect of taking into account the finite radius of
the interaction potential on the shape of the excitation spectrum is
weakened. At a certain density $\chi_0$ the parameter $\alpha$ in
the optical branch of the spectrum (\ref{67}) changes sign, becoming
positive at $\chi<\chi_0$. The value of $\chi_0$ is determined from
the condition that the numerator in the formula for $\tilde{\alpha}$
(\ref{70}) turns to zero. For the chosen above values of the
potential parameters and $x=1$ it is equal to $\chi_0\approx 0.1695$, %
which corresponds to the density $n_0=3.6\cdot\!10^{21}$\,cm$^{-3}$. %
For $x=0.1$ these values are equal to $\chi_0\approx 0.1004$, and %
$n_0=2.1\cdot\!10^{21}$\,cm$^{-3}$. The dispersion curves for the
density $n=10^{15}$\,cm$^{-3}$ close to that achieved in atomic
gases and the same parameters and the mass of particles as above are
shown in Fig.\,3.
\vspace{0mm} %
\begin{figure}[h!]
\vspace{-0mm}  \hspace{0mm}
\includegraphics[width = 8.0cm]{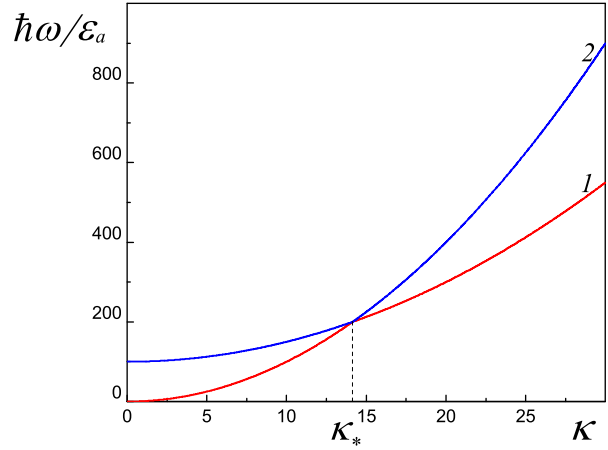} 
\vspace{-3mm} %
\caption{\label{fig03} 
Phonon (\!{\it 1}) $\varepsilon=\hbar\omega^{(-)}\!\big/\varepsilon_a$ %
and optical (\!{\it 2}) $\varepsilon=\hbar\omega^{(+)}\!\big/\varepsilon_a$ %
branches in the rarefied system with $n=10^{15}$\,cm$^{-3}$ and parameters: %
$m=6.65\cdot\!10^{-24}$\,g,\,\,$r_0=2.24\cdot\!10^{-8}$\,cm,\,\,$x=1$,\,\,$\chi=4.7\cdot\!10^{-8}$,\,\,$\theta=100$,\,\,$J=0.1$. %
}%
\end{figure}

The shape of this spectrum is close to that previously obtained in
\cite{R11} for the local interaction. At low densities the energy on
both branches increases monotonically and at $\kappa_*\approx 14.142$ %
the branches approach each other so strongly that, apparently, the
transformation of phonon excitations into optical ones and vice
versa becomes possible here. This question is of independent
interest, but is beyond the scope of this work.

\section{Dispersion of sound}\vspace{-2mm} %
In the limit $k\rightarrow 0$ for the phonon branch, the energy
depends linearly on the wave number $\varepsilon_k^{(-)}\!=\tilde{c}\kappa$. %
With increasing $k$ this dependence deviates from linearity. If the
deviation is towards lower energies then one speaks of a normal
dispersion, and if it is towards higher energies -- of an anomalous
dispersion. At very low frequencies the speed of sound is equal to
the speed of the hydrodynamic first sound. At higher frequencies
$\omega\gg\nu$, where $\nu$ is the collision frequency, there is a
transition to the zero-sound mode. The phase velocity of the zero
sound in quantum liquids turns out to be greater than the phase
velocity of the first sound \cite{R37}. It should also be noted that
the sound part of the spectrum is practically insensitive to the
superfluid transition.

Accounting for the deviation of the sound dispersion law from a
linear dependence is important, in particular, for studying kinetic
processes in superfluid $^4$He. In early works on the kinetics of
helium it was assumed that the deviation from a linear dependence
occurs in direction of decreasing energy (the normal dispersion) \cite{R12}. %
Subsequently, precision measurements of the initial part of the
dispersion curve revealed the deviation of the phase velocity of
excitations towards larger values (an anomalous dispersion) \cite{R19,R38}. %
The calculation carried out in this work also leads to an anomalous
dispersion. The phonon branch and its comparison with the sound
linear dependence are shown in Fig.\,4.
\vspace{0mm} %
\begin{figure}[h!]
\vspace{-6mm}  \hspace{0mm}
\includegraphics[width = 7.2cm]{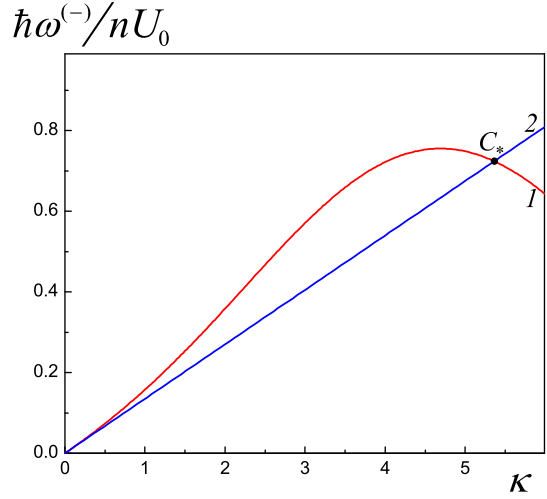} 
\vspace{-4mm} %
\caption{\label{fig04} 
Dispersion of phonons. Coordinates of the point of intersection of
the phonon branch with the linear dependence $C_*(5.36,\,0.724)$. %
}%
\end{figure}
\begin{figure}[h!]
\vspace{-6mm}  \hspace{0mm}
\includegraphics[width = 14.7cm]{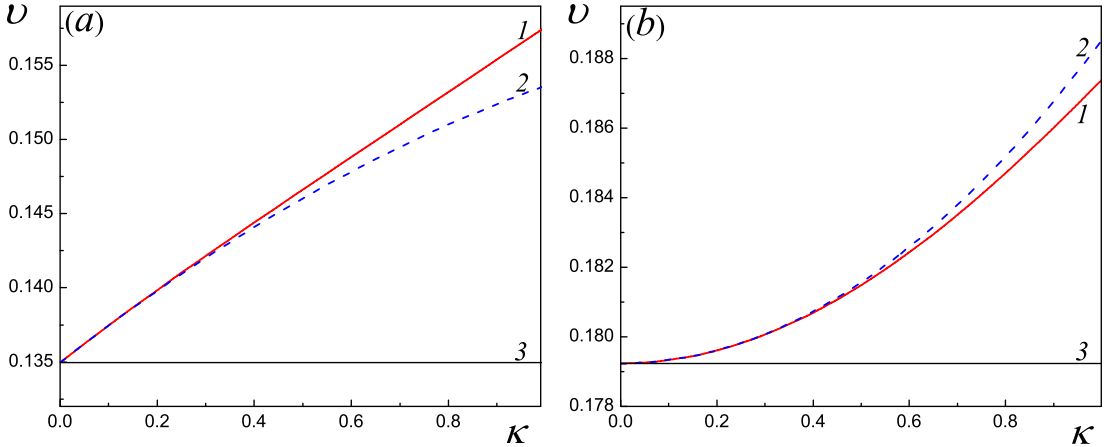} 
\vspace{-4mm} %
\caption{\label{fig05} 
(\!{\it 1}) The exact phase velocity $\upsilon_{ph}(\kappa)$; %
(\!{\it 2}) the phase velocity $\upsilon_{ph}^{(2)}(\kappa)$ according to formula (\ref{71}); %
(\!{\it 3}) the phase velocity without account of dispersion $\upsilon_{ph}^{(0)}$. %
(\!{\it a}) Calculation with parameters $\theta=100,\,J=0.1,\,x=0.1,\,\chi=1;\,\upsilon_{ph}^{(0)}=0.135$. %
(\!{\it b})~Calculation with parameters $\theta=100,\,J=0.0,\,x=0.1,\,\chi=1;\,\upsilon_{ph}^{(0)}=0.179$. %
}\vspace{-1mm} %
\end{figure}

As we can see, the phonon branch deviates from the linear dependence
towards higher energies and crosses the linear dependence at point
$C_*$. Using the expansions of coefficients given in Appendix C, for
the phase velocity to within a quadratic correction we obtain the
following formula:
\nopagebreak[4]\newpage$\vspace{-7mm}$ %
\begin{equation}\label{71}
\begin{array}{ccc}
\displaystyle{%
  \upsilon_{ph}(\kappa)\equiv\frac{\hbar}{nU_0r_0}\frac{\omega^{(-)}}{k}=  %
  \sqrt{\frac{\tilde{N}_2}{\tilde{L}_0}}\,+\frac{\tilde{N}_3}{2\sqrt{\tilde{L}_0\tilde{N}_2}}\,\kappa\,\, +  %
  \frac{\big(4\tilde{L}_0^2\tilde{N}_2\tilde{N}_4-4\tilde{L}_0\tilde{L}_2\tilde{N}_2^2+4\tilde{N}_2^3-\tilde{L}_0^2\tilde{N}_3^2\big)} %
       {8\tilde{L}_0^{5/2}\tilde{N}_2^{3/2}}\,\kappa^2.  %
}%
\end{array}
\end{equation}
Dependences of the phase velocity on the wave number are shown in
Fig.\,5. Here curve {\it 1} corresponds to the exact phase velocity
$\upsilon_{ph}(\kappa)$, curve {\it 2} corresponds to the phase
velocity calculated by formula (\ref{71}), and line {\it 3} -- to
the phase velocity in the absence of dispersion.

It draws attention that the linear contribution in the expression
for the phase velocity (\ref{71}) exists only when the attraction in
the interparticle potential is taken into account (Fig.\,5{\it a}).
In the model of ``semi-transparent sphere'', when there is no
attraction $(J=0)$, the linear term in (\ref{71}) vanishes (Fig.\,5{\it b}). %

\section{Discussion. Conclusions}\vspace{-2mm} %
In his first work on the theory of superfluid helium \cite{R39},
Landau suggested that there are two types of elementary excitations
in it -- the sound excitations and those that have a gap at zero
momentum. However, it turned out that the excitations with a gap
make an insufficient contribution for a correct description of
thermodynamic quantities. In this connection Landau proposed in
\cite{R40} his famous unified nonmonotonic dispersion curve. For
such law of dispersion the main contribution to thermodynamic
quantities comes from the gapless excitations at small momenta
(phonons) and the excitations with a large momentum near the minimum
of the dispersion curve (rotons).
\begin{figure}[b!]
\vspace{-3mm}  \hspace{0mm}
\includegraphics[width = 7cm]{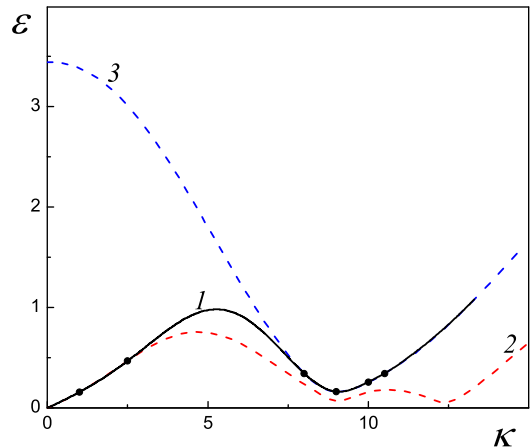} 
\vspace{-4mm} %
\caption{\label{fig06} 
Formation of the unified spectrum $\varepsilon_a(\kappa)$  (curve \!{\it 1}) %
from (\!{\it 2}) the phonon $\varepsilon_{ph}(\kappa)=\big(\hbar\omega^{(-)}\!\big/nU_0\big)$ %
and (\!{\it 3}) the optical-roton $\varepsilon_{o\!p}(\kappa)=\big(\hbar\omega^{(+)}\!\big/nU_0\big)$ %
branches. Calculation parameters: $n=2.2\cdot\!10^{22}$\,cm$^{-3}$, %
$m=6.65\cdot\!10^{-24}$\,g, $r_0=2.24\cdot\!10^{-8}$\,cm,\, %
$x=0.1,\,\chi=1,\,\theta=100,\,J=0.1$. %
}\vspace{0mm} %
\end{figure}

The shape of the curve in helium was experimentally studied and
analyzed in experiments on neutron scattering \cite{R13,R14,R15,R16,R17,R18,R19,R20}. %
These experiments point to the complex structure of the spectrum.
The scattering peaks observed in experiments can be characterized by
their total intensity $Z(q)$ and width at half maximum $W(q)$. These
characteristics behave differently in different regions of change of
the transmitted wave number $q$. A sharp peak is observed in the
region $0<q<0.65\,\textrm{{\AA}}^{-1}$, corresponding to the weakly
damped sound excitations. Upon transition to the maxon-roton region
$q>0.65\,\textrm{{\AA}}^{-1}$ the behavior of the characteristics
$Z(q)$ and $W(q)$ changes significantly, which indicates the
probable different nature of the sound and maxon-roton regions of
the dispersion curve.

The scattering peak has the most complex structure in the transition
region \cite{R17,R18,R19,R20}. Here the sound component gradually
becomes strongly damped, but a narrow component appears, associated
with the maxon-roton excitations. In addition, apparently, there is
also a broad component. When processing the experimental data of
experiments \cite{R17,R18,R19,R20}, it was possible to distinguish
two components in the neutron scattering peak in the maxon-roton
region -- narrow and wide. At low temperature the narrow peak lies
somewhat higher than the broad peak \cite{R17}.

The calculations of dispersion curves performed in this work and the
results of experiments on neutron scattering \cite{R13,R14,R15,R16,R17,R18,R19,R20} %
give grounds to assume that the Landau curve consists of two parts
that belong to two different excitation branches. The initial region
of the spectrum is determined by the long-lived phonon excitations,
which become strongly damped at large wave numbers. Well-defined
long-lived excitations in the maxon-roton region of the spectrum are
a part of the optical excitation branch. Since in the theoretical
model considered in this work there were made a number of
significant approximations, which were mentioned above, then the
constructed dispersion curves are not accurate. Nevertheless, the
obtained results allow to propose a qualitative picture of formation
of the unified spectrum from two different excitation branches,
which is shown in Fig.\,6.

The interpolation curve in Fig.\,6 is given by the formula
\begin{equation} \label{72}
\begin{array}{ccc}
\displaystyle{%
  \varepsilon_a(\kappa)=\tilde{c}\kappa+\kappa\,\psi(\kappa), %
}%
\end{array}
\end{equation}
where $\psi(\kappa)$ is a polynomial constructed using the Lagrange
interpolation formula on the basis of several points on the phonon
and optical branches shown in Fig.\,6. It should be noted that
excitations on the optical branch at small momenta are difficult to
detect experimentally, since they make a negligibly small
contribution to thermodynamic quantities, and it is technically
difficult to measure excitations with small momenta in neutron
scattering experiments.

The main results of the work are the following: ({\it a}) The system
of equations for the single-particle condensate and pair
correlations is obtained with taking into account the finite radius
of the interaction potential between particles. ({\it b}) The
spectrum of elementary excitations is studied and it is shown that
it has two branches -- phonon and optical. At high densities the
dispersion curves are nonmonotonic and have the roton-like minima.
({\it c}) It is shown that the phonon spectrum has an anomalous
dispersion. ({\it d}) On the basis of the performed calculations and
analysis of experiments on inelastic neutron scattering, an
assumption is made about the complex nature of the Landau spectrum
in superfluid $^4$He.

\appendix

\section{General form of coefficients in the dispersion equation (55)} %
\vspace{-8mm}
\begin{equation} \nonumber
\begin{array}{ccc}
\displaystyle{%
  L_k\equiv A_k+C_k,\qquad   N_k\equiv A_kC_k-B_kD_k,   %
}%
\end{array}
\end{equation}
\vspace{-6mm}
\begin{equation}\label{A01}
\begin{array}{ccc}
\displaystyle{%
  A_k=\Big[\varepsilon_k^{(1)}+2xnU_0\Big]\!\Big[\varepsilon_k^{(2)}-2nU_0(x+z)\Big] +  %
}\vspace{2mm}\\ %
\displaystyle{\hspace{10mm}%
  +\,2xn\big(U_0+\Delta U_{k/2}\big)\big[2nU_0(2w-x)+\alpha_k\big] + %
}\vspace{2mm}\\ %
\displaystyle{\hspace{22mm}%
  +\,2xn\big[2U_0+\big(\Delta U_k+\Delta U_{k/2}\big)\big]\!\big[2nU_0(x+z)-\alpha_k\big], %
}%
\end{array}
\end{equation}
\vspace{-3mm} %
\begin{equation}\label{A02}
\begin{array}{ccc}
\displaystyle{%
  B_k=xn\big(U_0+\Delta U_{k/2}\big)\Big[\varepsilon_k^{(1)}+\,\varepsilon_k^{(3)}+2nU_0(2w-z-2x)-2xn\big(\Delta U_k+\Delta U_{k/2}\big)\Big],  %
}%
\end{array}
\end{equation}
\vspace{-5mm} %
\begin{equation}\label{A03}
\begin{array}{ccc}
\displaystyle{%
  C_k=\Big[\varepsilon_k^{(3)}+2nU_0(2w-z-x)\Big]^{\!2} +  %
  2xn\big(U_0+\Delta U_{k/2}\big)\big[2nU_0(2w-z-3x)+\gamma_k-2\delta_k\big], %
}%
\end{array}
\end{equation}
\vspace{-5mm} %
\begin{equation}\label{A04}
\begin{array}{ccc}
\displaystyle{%
  D_k=2\Big[\varepsilon_k^{(3)}+2nU_0(2w-z-x)\Big]\!\big[2nU_0(2w-x)+\alpha_k\big]\,+
}\vspace{2mm}\\ %
\displaystyle{\hspace{5mm}%
     +2\Big[\varepsilon_k^{(2)}-2nU_0(x+z)\Big]\!\big[2nU_0(2w+2z+x)+\gamma_k\big]\,+
}\vspace{2mm}\\ %
\displaystyle{\hspace{-6mm}%
     +\,4\big[nU_0(4x+3z)+\delta_k\big]\!\big[2nU_0(x+z)-\alpha_k\big].
}%
\end{array}
\end{equation}
In formulas (A1)\,--\,(A4) the notation is used:
\begin{equation}\label{A05}
\begin{array}{ccc}
\displaystyle{%
  \alpha_k\equiv n\Delta U_k(1-2x-z), \quad \beta_k\equiv n\Delta U_k(1+2x+z), \quad \gamma_k\equiv n\Delta U_k(1+2x+3z),
}\vspace{3mm}\\ %
\displaystyle{\hspace{5mm}%
  \delta_k\equiv n\Delta U_k(x+z) + n\Delta U_{k/2}(3x+2z),
}\vspace{3mm}\\ %
\displaystyle{\hspace{-6mm}%
  \varepsilon_k^{(1)}\equiv\varepsilon_k+\beta_k, \quad  \varepsilon_k^{(2)}\equiv\varepsilon_k+\alpha_k, \quad
  \varepsilon_k^{(3)}\equiv\frac{\varepsilon_k}{2}+2n\Delta U_{k/2}(2-x).
}%
\end{array}
\end{equation}

\section{Coefficients of the dispersion equation (55) for the potential (26)} %
\vspace{-8mm} %
\begin{equation}\label{B01}
\begin{array}{ccc}
\displaystyle{%
  \frac{A_k}{n^2U_0^2}\equiv \tilde{A}_k=4x(z+2w)+2b\Big[-\zeta z\kappa^2-3(x+z)(1+z)f(\kappa)+6x(z+2w)f\Big(\frac{\kappa}{2}\Big)\Big]\, + %
}\vspace{2mm}\\ %
\displaystyle{\hspace{0mm}%
  + \,b^2\Big[\zeta^2\kappa^4 + 6\zeta\kappa^2f(\kappa)+9(1+z)(1-2x-z)f^2(\kappa)\Big], %
}%
\end{array}
\end{equation}
\vspace{-4mm} %
\begin{equation}\label{B02}
\begin{array}{ccc}
\displaystyle{%
  \frac{B_k}{n^2U_0^2}\equiv \tilde{B}_k=2x(2w-2x-z)+3bx\!\left[\frac{1}{2}\zeta\kappa^2+(1+z)f(\kappa)+2(2+2w-4x-z)f\Big(\frac{\kappa}{2}\Big)\right]\, + %
}\vspace{2mm}\\ %
\displaystyle{\hspace{0mm}%
  + \,9b^2f\Big(\frac{\kappa}{2}\Big)x\!\left[\frac{1}{2}\zeta\kappa^2+(1+z)f(\kappa)+4(1-x)f\Big(\frac{\kappa}{2}\Big)\right], %
}%
\end{array}
\end{equation}
\vspace{-4mm} %
\begin{equation}\label{B03}
\begin{array}{ccc}
\displaystyle{%
  \frac{C_k}{n^2U_0^2}\equiv \tilde{C}_k=4\big(4w^2+2xw-4x^2-2wz-xz\big)\, + %
}\vspace{2mm}\\ %
\displaystyle{\hspace{0mm}%
  +\,2b\!\left[(2w-x-z)\zeta\kappa^2+3x(1+z)f(\kappa)+6\big(8w-4x-2xw-4x^2-4z-xz\big)f\Big(\frac{\kappa}{2}\Big)\right] %
}\vspace{2mm}\\ %
\displaystyle{\hspace{0mm}%
  + \,b^2\left[\frac{\zeta^2}{4}\kappa^4+6(2-x)\zeta\kappa^2f\Big(\frac{\kappa}{2}\Big)+72\big(2-2x-x^2-xz\big)f^2\Big(\frac{\kappa}{2}\Big) %
  + 18x(1+z)f(\kappa)f\Big(\frac{\kappa}{2}\Big)\right], %
}%
\end{array}
\end{equation}
\vspace{-4mm} %
\begin{equation}\label{B04}
\begin{array}{ccc}
\displaystyle{%
  \frac{D_k}{n^2U_0^2}\equiv \tilde{D}_k=8\big(4w^2+2x^2-2xw-2wz+3xz\big)\, + %
}\vspace{2mm}\\ %
\displaystyle{\hspace{0mm}%
  +\,4b\Big[\Big(3w+2z+\frac{x}{2}\Big)\zeta\kappa^2+3\big(4w+6x^2+5xz-4xw-2wz-5x-3z\big)f(\kappa)\,+ %
}\vspace{2mm}\\ %
\displaystyle{\hspace{0mm}%
  +\,6\big(4w-2x+6xw+4wz+xz\big)f\Big(\frac{\kappa}{2}\Big)\Big]+ %
}\vspace{2mm}\\ %
\displaystyle{\hspace{0mm}%
  + \,3b^2\!\left[(3+2x+5z)\zeta\kappa^2f(\kappa)+6(1+z)(1-2x-z)f^2(\kappa) %
  + 24(1-2x-z)^2f(\kappa)f\Big(\frac{\kappa}{2}\Big)\right], %
}%
\end{array}
\end{equation}
\vspace{-4mm} %
\begin{equation}\label{B05}
\begin{array}{ccc}
\displaystyle{%
  L_k=A_k+C_k=(nU_0)^2\tilde{L}_k, \qquad \tilde{L}_k=\tilde{A}_k + \tilde{C}_k, %
}\vspace{2mm}\\ %
\displaystyle{\hspace{0mm}%
  N_k=A_kC_k-B_kD_k=(nU_0)^4\tilde{N}_k, \qquad \tilde{N}_k=\tilde{A}_k \tilde{C}_k-\tilde{B}_k\tilde{D}_k. %
}%
\end{array}
\end{equation}
The notation is used:
\begin{equation}\label{B06}
\begin{array}{ccc}
\displaystyle{%
  \zeta\equiv\frac{\varepsilon_a}{\chi I}=\frac{1}{\theta\chi}, \qquad %
  b=\frac{1}{1-J}, \qquad \kappa\equiv r_0k.
}%
\end{array}
\end{equation}
The function $f(\kappa)$ is defined by formulas (\ref{60}),\,(\ref{61}).

\section{Coefficients of the dispersion equation (55) for the potential (26) \newline in the long-wavelength limit $\kappa\ll 1$} %
\vspace{-8mm} %
\begin{equation}\label{C01}
\begin{array}{ccc}
\displaystyle{%
  \tilde{A}_k\approx \tilde{A}_0+\tilde{A}_2\kappa^2+\tilde{A}_3\kappa^3+\tilde{A}_4\kappa^4, \qquad %
  \tilde{B}_k\approx \tilde{B}_0+\tilde{B}_2\kappa^2+\tilde{B}_3\kappa^3+\tilde{B}_4\kappa^4, %
}\vspace{2mm}\\ %
\displaystyle{\hspace{0mm}%
  \tilde{C}_k\approx \tilde{C}_0+\tilde{C}_2\kappa^2+\tilde{C}_3\kappa^3+\tilde{C}_4\kappa^4, \qquad %
  \tilde{D}_k\approx \tilde{D}_0+\tilde{D}_2\kappa^2+\tilde{D}_3\kappa^3+\tilde{D}_4\kappa^4, %
}\vspace{2mm}\\ %
\displaystyle{\hspace{0mm}%
  \tilde{L}_k\approx \tilde{L}_0+\tilde{L}_2\kappa^2+\tilde{L}_3\kappa^3+\tilde{L}_4\kappa^4,  %
}\vspace{2mm}\\ %
\displaystyle{\hspace{0mm}%
  \tilde{L}_0\equiv \tilde{A}_0+\tilde{C}_0, \quad \tilde{L}_2\equiv \tilde{A}_2+\tilde{C}_2, \quad  %
  \tilde{L}_3\equiv \tilde{A}_3+\tilde{C}_3, \quad \tilde{L}_4\equiv \tilde{A}_4+\tilde{C}_4,        %
}\vspace{2mm}\\ %
\displaystyle{\hspace{0mm}%
  \tilde{N}_k\approx \tilde{N}_2\kappa^2+\tilde{N}_3\kappa^3+\tilde{N}_4\kappa^4,  %
}\vspace{2mm}\\ %
\displaystyle{\hspace{0mm}%
  \tilde{N}_2\equiv \tilde{A}_0\tilde{C}_2+\tilde{C}_0\tilde{A}_2-\tilde{B}_0\tilde{D}_2-\tilde{D}_0\tilde{B}_2,  %
}\vspace{2mm}\\ %
\displaystyle{\hspace{0mm}%
  \tilde{N}_3\equiv \tilde{A}_0\tilde{C}_3+\tilde{C}_0\tilde{A}_3-\tilde{B}_0\tilde{D}_3-\tilde{D}_0\tilde{B}_3,  %
}\vspace{2mm}\\ %
\displaystyle{\hspace{0mm}%
  \tilde{N}_4\equiv \tilde{A}_0\tilde{C}_4+\tilde{C}_0\tilde{A}_4+\tilde{A}_2\tilde{C}_2-\tilde{B}_0\tilde{D}_4-\tilde{D}_0\tilde{B}_4-\tilde{B}_2\tilde{D}_2.  %
}\vspace{0mm} 
\end{array}
\end{equation}
Here\vspace{-1mm} %
\begin{equation}\label{C02}
\begin{array}{ccc}
\displaystyle{%
  \tilde{A}_0=4x(2w+z), \qquad \tilde{B}_0=2x(2w-z-2x), %
}\vspace{2mm}\\ %
\displaystyle{\hspace{0mm}%
  \tilde{C}_0=4\big(\!-4x^2+4w^2+2xw-2wz-xz\big), %
}\vspace{2mm}\\ %
\displaystyle{\hspace{0mm}%
  \tilde{D}_0=8\big(\!2x^2+4w^2-2xw-2wz+3xz\big),  %
}%
\end{array}
\end{equation}
\vspace{-3mm} %
\begin{equation}\label{C03}
\begin{array}{ccc}
\displaystyle{%
  \tilde{A}_2=-2zb\zeta + u\big(2z^2+xz-2xw+2x+2z\big), %
}\vspace{2mm}\\ %
\displaystyle{\hspace{0mm}%
  \tilde{B}_2=\frac{x}{2}\big[3b\zeta-u(4+z+2w-4x)\big], %
}\vspace{2mm}\\ %
\displaystyle{\hspace{0mm}%
  \tilde{C}_2=2(2w-x-z)b\zeta + u\big(2x-8w+4z+2xw-xz+4x^2\big),  %
}\vspace{2mm}\\ %
\displaystyle{\hspace{0mm}%
  \tilde{D}_2=4\!\left(3w+2z+\frac{x}{2}\right)\!b\zeta + %
              4u\Big(6x-6w+3z+xw-6x^2-\frac{11}{2}xz\Big), %
}%
\end{array}
\end{equation}
\vspace{-3mm} %
\begin{equation}\label{C04}
\begin{array}{ccc}
\displaystyle{%
  \tilde{A}_3=\frac{\pi}{32}(b-1)\big(4x+4z+3xz-2xw+4z^2\big), %
}\vspace{2mm}\\ %
\displaystyle{\hspace{0mm}%
  \tilde{B}_3=-\frac{\pi}{64}(b-1)\,x(6+3z+2w-4x), %
}\vspace{2mm}\\ %
\displaystyle{\hspace{0mm}%
  \tilde{C}_3=-\frac{\pi}{32}(b-1)\big(8w-4z+3xz-2xw-4x^2\big),  %
}\vspace{2mm}\\ %
\displaystyle{\hspace{0mm}%
  \tilde{D}_3=-\frac{\pi}{16}(b-1)\big(24x^2 -10xw + 21xz -4wz - 22x +20w -12z \big),  %
}%
\end{array}
\end{equation}
\vspace{-3mm} %
\begin{equation}\label{C05}
\begin{array}{ccc}
\displaystyle{%
  \tilde{A}_4=-6bb_4\Big(x+z+z^2+\frac{7}{8}xz-\frac{1}{4}xw\Big) + b^2\zeta^2 - 2b\zeta u + u^2\big(1-2x-2xz-z^2\big), %
}\vspace{2mm}\\ %
\displaystyle{%
  \tilde{B}_4=\frac{3}{8}bb_4x(10+7z-4x+2w) -\frac{3}{8}ub\zeta x + \frac{u^2}{4}x(2+z-x), %
}\vspace{2mm}\\ %
\displaystyle{\hspace{0mm}%
  \tilde{C}_4= \frac{1}{4}\Big(b^2\zeta^2-2u(2-x)b\zeta + 2u^2\big(2-x-x^2\big)+\frac{3}{4}bb_4\big(4x+8w-4z+7xz-2xw-4x^2\big), %
}\vspace{2mm}\\ %
\displaystyle{\hspace{0mm}%
  \tilde{D}_4=\frac{3}{2}bb_4\big(48x^2+41xz-26xw-12wz-42x+36w-24z\big) +  %
}\vspace{2mm}\\ %
\displaystyle{\hspace{0mm}%
  +\,4u^2\big(1-3x-z+2x^2+xz\big)-ub\zeta(3+2x+5z).  %
}%
\end{array}
\end{equation}
The notation is used:
\begin{equation}\label{C06}
\begin{array}{ccc}
\displaystyle{%
  u\equiv\frac{(1/5-J)}{2(1-J)}=\frac{1}{2}\!\left(\!1-\frac{4}{5}\,b\right), \qquad %
  b_4\equiv\frac{1}{120}\!\left(\frac{1}{7}+J\right). %
}%
\end{array}
\end{equation}


\end{document}